\documentclass{aa}
\usepackage{txfonts}
\usepackage{graphicx}
\usepackage{natbib}
\usepackage{amssymb}
\usepackage{lscape}
\bibpunct{(}{)}{;}{a}{}{,}

\def\chandra{{\it Chandra~}}

\def\swift{{\it Swift~}}

\def\m31{{M~31}}
\def\msun{{$M_{\sun}$}}

\def\md{HPH2014~}
\def\mdk{HPH2014}

\newcommand{\nh}{\hbox{$N_{\rm H}$}~}

\newcommand{\hcm}[1]{$\times 10^{#1}$ cm$^{-2}$}

\newcommand{\tpower}[1]{$\times 10^{#1}$}
\newcommand{\power}[1]{$10^{#1}$}

\newcommand{\ton}{$t_{\mbox{\small{on}}}$~}
\newcommand{\toff}{$t_{\mbox{\small{off}}}$~}

\newcommand{\ttwo}{$t_{2,R}$~}

\newcommand{\tonk}{$t_{\mbox{\small{on}}}$}
\newcommand{\toffk}{$t_{\mbox{\small{off}}}$}

\newcommand{\eton}{t_{\mbox{\small{on}}}}
\newcommand{\etoff}{t_{\mbox{\small{off}}}}

\newcommand{\emej}{M_{\mbox{\small{ej,H}}}}

\def\nova{{M31N~2008-12a~}}
\def\novak{{M31N~2008-12a}}

\def\dwb{{DWB2014~}}
\def\dwbk{{DWB2014}}

\def\hnd{{HND2014~}}
\def\hndk{{HND2014}}

\def\opt{{DHS2015~}}
\def\optk{{DHS2015}}

\begin{document}

\title{A remarkable recurrent nova in \m31: The predicted 2014 outburst in X-rays with \swift}

\author{M.~Henze\inst{1}
	\and J.-U.~Ness\inst{1}
  \and M.~J.~Darnley\inst{2}
	\and M.~F.~Bode\inst{2}
  \and S.~C.~Williams\inst{2}
  \and A.~W.~Shafter\inst{3}
  \and G.~Sala\inst{4,5}
  \and M.~Kato\inst{6}
  \and I.~Hachisu\inst{7}
  \and M.~Hernanz\inst{8}
}

\institute{European Space Astronomy Centre, P.O. Box 78, 28692 Villanueva de la Ca\~{n}ada, Madrid, Spain\\
	email: \texttt{mhenze@sciops.esa.int}
	\and Astrophysics Research Institute, Liverpool John Moores University, IC2 Liverpool Science Park, Liverpool, L3 5RF, UK
  \and Department of Astronomy, San Diego State University, San Diego, CA 92182, USA
  \and Departament de F\'isica i Enginyeria Nuclear, EUETIB, Universitat Polit\`ecnica de Catalunya, c/ Comte d'Urgell 187, 08036 Barcelona, Spain
  \and Institut d'Estudis Espacials de Catalunya, c/Gran Capit\`a 2-4, Ed. Nexus-201, 08034, Barcelona, Spain
  \and Department of Astronomy, Keio University, Hiyoshi, Yokohama 223-8521, Japan
  \and Department of Earth Science and Astronomy, College of Arts and Sciences, The University of Tokyo, Komaba, Meguro-ku, Tokyo 153-8902, Japan
  \and Institut de Ci\`encies de l'Espai (CSIC-IEEC), Campus UAB, C/Can Magrans s/n, E-08193 Cerdanyola del Valles, Spain
}

\date{Received 5 March 2015 / Accepted 23 April 2015}

\abstract
{The \m31 nova \nova was recently found to be a recurrent nova (RN) with a recurrence time of about 1 year. This is by far the fastest recurrence time scale of any known RN.}
{Our optical monitoring programme detected the predicted 2014 outburst of \nova in early October. We immediately initiated an X-ray/UV monitoring campaign with \swift to study the multiwavelength evolution of the outburst.}
{We monitored \nova with daily \swift observations for 20 days after discovery, covering the entire supersoft X-ray source (SSS) phase.}
{We detected SSS emission around day six after outburst. The SSS state lasted for approximately two weeks until about day 19. \nova was a bright X-ray source with a high blackbody temperature.}
{The X-ray properties of this outburst were very similar to the 2013 eruption. Combined X-ray spectra show a fast rise and decline of the effective blackbody temperature. The short-term X-ray light curve showed strong, aperiodic variability which decreased significantly after about day 14. Overall, the X-ray properties of \nova are consistent with the average population properties of \m31 novae. The optical and X-ray light curves can be scaled uniformly to show similar time scales as those of the Galactic RNe U~Sco or RS~Oph. The SSS evolution time scales and effective temperatures are consistent with a high-mass WD. We predict the next outburst of \nova to occur in autumn 2015.}


\titlerunning{RN \novak: 2014 outburst in X-rays}

\maketitle

%
%
\section{Introduction}
\label{sec:intro}
%
Novae are the product of powerful outbursts occurring on white dwarfs (WD) in close binary systems where hydrogen-rich material accreted from the companion star accumulates on the WD surface until hydrogen fusion in degenerate matter leads to an explosive ejection of the accreted envelope. The expanding hot pseudo photosphere rapidly increases the optical luminosity of the system by four to eight orders of magnitude before fading more slowly. This ``new star'' is the optical nova. For recent reviews on nova science see \citet{2008clno.book.....B}.

When the ejected envelope expands, deeper and hotter layers of the pseudo photosphere become visible. This leads to a hardening of the nova spectrum at an approximately constant bolometric luminosity until ultimately a supersoft X-ray source (SSS) emerges \citep[e.g.][]{2006ApJS..167...59H,2008ASPC..401..139K}. Powered by stable hydrogen burning, the SSS phase typically lasts from months to years. Its end indicates the cessation of the burning and the descent of the WD luminosity and temperature back to quiescence.

The underlying WD is not significantly affected by a single nova outburst. After a certain time, accretion resumes and leads to the next outburst. Recurrent novae (RNe), in contrast to classical novae (CNe), are those systems for which more than one outburst has been observed. This is a phenomenological definition based on approximately a century of modern astronomical observations. All novae can show repeated outbursts \citep{2007MNRAS.374.1449E} on typical time scales of megayears down to a few months for the most extreme objects \citep{2005ApJ...623..398Y,2014ApJ...793..136K}. Together with an infrequent observational coverage these different time scales mean that currently only ten Galactic RNe are known \citep{2010ApJS..187..275S}. However, \citet{2014ApJ...788..164P} recently suggested that about 25\% of the $\sim400$ known Galactic novae might be potential RNe. Recurrent novae are good candidates for type-Ia supernova progenitors \citep[e.g.][]{2012BASI...40..393K}.

For our large neighbour galaxy \m31, \citet{2015ApJS..216...34S} recently reported a comprehensive archival study revealing 16 likely RNe among the almost one thousand known outbursts \citep[see the online catalogue\footnote{http://www.mpe.mpg.de/$\sim$m31novae/opt/m31/index.php} of][]{2007A&A...465..375P}. In X-rays, \citet[][hereafter \mdk]{2014A&A...563A...2H} presented updated results from a dedicated monitoring survey of \m31 novae which, together with the archival analysis of \citet{2005A&A...442..879P,2007A&A...465..375P}, resulted in an unprecedentedly large sample of 79 novae with SSS counterpart \citep[see also][]{2010A&A...523A..89H,2011A&A...533A..52H}. A statistical analysis of the sample showed strong correlations between X-ray and optical nova parameters (\mdk). Its proximity and size make \m31 the ideal target for extragalactic nova surveys.

In late 2012 we realised that the repeated detections of \nova established the object as a likely RN \citep{2012ATel.4503....1S}. Subsequently, the 2013 eruption was monitored closely in X-rays by \citet[][hereafter \hndk]{2014A&A...563L...8H} and studied by \citet[][hereafter \dwbk]{2014A&A...563L...9D} in the optical \citep[see also][]{2014ApJ...786...61T}. The detection of a bright and hot SSS phase clearly showed that another nova outburst had occurred only approximately one year after the previous outburst.

Earlier outbursts had been detected in 2008 (the eponymous M31N 2008-12a found by Kabashima \& Nishiyama\footnote{http://www.cbat.eps.harvard.edu/CBAT\_M31.html\#2008-12a}) and 2011 (M31N 2011-10e; Korotkiy \& Elenin\footnote{http://www.cbat.eps.harvard.edu/unconf/followups/\\J00452885+4154094.html}), albeit only with relatively sparse single-filter photometry. \citet{2014ApJ...786...61T} found another outburst in archival data of Dec 2009. Additionally, in \hnd we recall two X-ray detections with ROSAT in 1992 and 1993 \citep{1995ApJ...445L.125W} and one \chandra detection in 2001 \citep{2004ApJ...609..735W} that can now be attributed with high probability to earlier outbursts of \novak.

The sixth outburst in seven years of nova \nova was predicted in \hndk/\dwb and successfully discovered in early October 2014 by \citet{2014ATel.6527....1D} as the result of a dedicated optical monitoring programme \citep[see also][]{2014ATel.6535....1D}. Optical spectroscopy confirmed the nova outburst and described a spectrum consistent with the previous eruptions \citep{2014ATel.6540....1D}. The optical and UV properties of the 2014 outburst are described in detail in \citet[][hereafter \optk]{opt}.

We initiated a \swift monitoring campaign that detected the SSS phase soon after the outburst \citep{2014ATel.6558....1H,2014ATel.6565....1H} and followed the X-ray light curve until the source had disappeared \citep{2014ATel.6604....1H}. This paper describes the SSS evolution of \nova during the 2014 outburst and draws comparisons to the 2013 X-ray data.

%
\section{Observations and data analysis}
\label{sec:obs}
%
Immediately after the 2014 outburst of \nova was announced by \citet{2014ATel.6527....1D}, we initiated a daily monitoring campaign with \swift \citep{2004ApJ...611.1005G}. This programme began only 18 hours after the discovery and was designed to cover the complete SSS phase, as predicted, based on the 2013 outburst (see \hndk). In 2013 the SSS was already visible at the time of our first X-ray observation \citep[on day 6 after the eruption, see \hndk,][]{2013ATel.5627....1H}. This time, we were determined to constrain the SSS turn on of the nova more accurately. All monitoring observations are summarised in Table\,\ref{tab:obs}. We were granted a 20-day monitoring with daily observations of 8 ks duration whenever the scheduling permitted. This monitoring, with exposure times that were considerably longer than in 2013, was designed to study the (spectral) variability of \nova in greater detail.

The \swift X-ray telescope \citep[XRT;][]{2005SSRv..120..165B} data analysis started from the cleaned event files as produced at the Swift Data Center using HEASoft version 6.13. In our analysis we used HEASoft version 6.16. Count rates and upper limits were estimated with the XIMAGE (version 4.5.1) source statistics (\texttt{sosta}) tool. The background level was derived from a source-free region near the object. All estimates include corrections for vignetting, dead time, and the point spread function (PSF) of the source based on the merged data from all detections. All uncertainties correspond to 1$\sigma$ confidence and all upper limits to 3$\sigma$ confidence unless otherwise noted.

The X-ray spectra were extracted using the Xselect software (version 2.4c) and analysed using XSPEC \citep[][version 12.8.2]{1996ASPC..101...17A}. Our XSPEC models assumed the ISM abundances from \citet{2000ApJ...542..914W}, the T\"ubingen-Boulder ISM absorption model (\texttt{TBabs} in XSPEC), and the photoelectric absorption cross-sections from \citet{1992ApJ...400..699B}. Spectra were binned to include at least one count per bin and fitted in XSPEC assuming Poisson statistics according to \citet{1979ApJ...228..939C}.

A typical \swift observation comprises several individual ``snapshots'' due to the low-earth orbit of the satellite. In this paper, we determined count rates and spectra to study the short-time (spectral) variability. This analysis was based on good-time interval (GTI) filtering of the event files. We created exposure maps for every snapshot and used identical source and background regions to extract source counts. The resulting observation parameters and measurements are listed in Table\,\ref{tab:obs_split}. For the variability analysis in Sect.\,\ref{sec:results} we excluded snapshots with exposure times of less than 50 s, because of the large uncertainties. All statistical tests on the X-ray variability were performed using the R software \citep{R_manual}.

We also analysed the \swift UV/optical telescope \citep[UVOT,][]{2005SSRv..120...95R} data and estimated the magnitude of the nova using the \texttt{uvotsource} tool. All magnitudes assume the UVOT photometric system \citep{2008MNRAS.383..627P} and have not been corrected for extinction. 

\section{Results}
\label{sec:results}
%
The SSS phase of nova \nova was detected by \swift to have started on day six after the 2014 outburst and lasted until day 18. The position, spectrum and light curve of the X-ray source are in excellent agreement with the 2013 measurements. This shows beyond reasonable doubt that we did detect supersoft X-ray emission from the 2014 outburst of \novak.

\subsection{X-ray light curve}
\label{sec:res_lc}
To determine the X-ray time scales more precisely we assume here that the optical outburst occurred on 2014-10-02.69 UT (MJD 56932.69) with an uncertainty of 0.21 d. This date is halfway between the discovery \citep[2014-10-02.90 UT; see][]{2014ATel.6527....1D} and the last non detection published by the iPTF collaboration \citep[2014-10-02.47 UT][]{2014ATel.6532....1C}. All time scales in this paper assume MJD 56932.69 as $t = 0$. Note, that \opt define their $t = 0$ reference date as the optical flux maximum, which occurred almost exactly one day after our assumed outburst (MJD 56933.7, see \optk).

\begin{figure}[t!]
  \resizebox{\hsize}{!}{\includegraphics[angle=0]{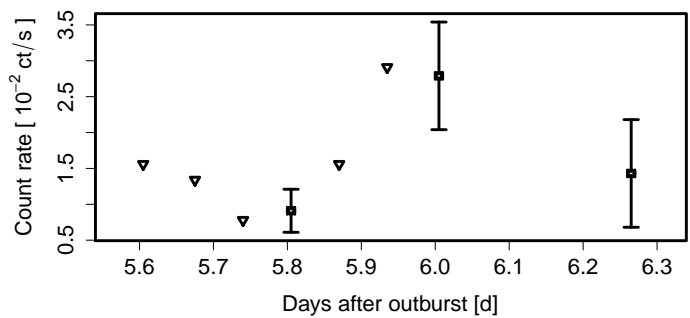}}
  \caption{\swift XRT light curve (0.2 - 10 keV) of nova \nova during day six after outburst. Triangles indicate upper limits. Based on the individual snapshots of observation 00032613048 (see Table\,\ref{tab:obs_split}). This observation likely shows the emergence of the SSS.}
  \label{fig:lc_onset}
\end{figure}

The light curve around day six, as shown in Fig.\,\ref{fig:lc_onset} (see also Table\,\ref{tab:obs_split}), suggests that we witnessed the gradual emergence of the SSS during this observation \citep[\swift ObsID 00032613048; see Table\,\ref{tab:obs} and also][]{2014ATel.6558....1H}. We conclude that the SSS turn-on time (\tonk) was 5.9 days after outburst (MJD = 56938.6; the midpoint of the observation on day six). We derive an uncertainty of 0.5 d, combined from the half duration of the XRT observation and the uncertainty of the optical outburst: $\eton =  (5.9 \pm 0.5)$~d.

\begin{figure}[t!]
  \resizebox{\hsize}{!}{\includegraphics[angle=0]{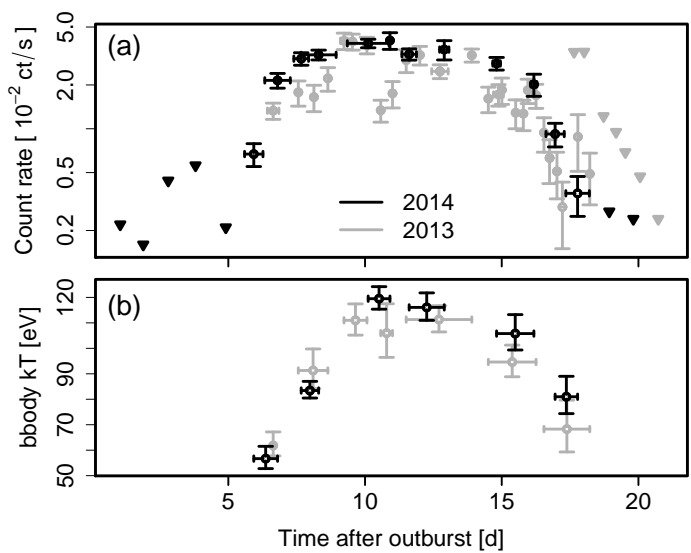}}
  \caption{\swift XRT (a) count rate light curve (0.2 - 10 keV) and (b) effective black-body temperature evolution of \nova during the 2014 outburst (black). In grey we show the corresponding data of the 2013 outburst (based on \hndk). Time is in days after 2014-10-02.69 UT (see Sect.\,\ref{sec:res_lc}) for the 2014 data. The error bars in time represent either (a) the duration of the observation or (b) the time between sets of observations (see Sect.\,\ref{sec:res_spec}). \textit{Panel a:} Triangles indicate upper limits. \textit{Panel b:} Sets of observations with similar spectra have been fitted simultaneously assuming a fixed \nh = $1.4$ \hcm{21}.}
  \label{fig:lc}
\end{figure}

The X-ray luminosity of the nova rose quickly until reaching a plateau around day seven to eight after outburst. The full \swift XRT light curve of the 2014 outburst is shown in Fig.\,\ref{fig:lc}a. This figure also shows the 2013 light curve from \hnd as a comparison in grey. Both light curves are very similar. We discuss these similarities in Sect.\,\ref{sec:disc_2013}.

After day 15, the X-ray flux started to decline markedly and the SSS had disappeared within a few days. We estimate the SSS turn off (\toffk) to have occurred between the end of observation 00032613060 (effectively on MJD 56950.84, the end of the fifth snapshot, see Table\,\ref{tab:obs_split}) and the start of the following observation 00032613061 (MJD 56951.42, see Table\,\ref{tab:obs}). Including the uncertainty on the optical outburst data this gives a $\etoff = (18.4 \pm 0.5)$~d.

\subsection{X-ray spectrum}
\label{sec:res_spec}
In Fig.\,\ref{fig:lc}b we show the evolution of the effective temperature of \nova during the SSS phase. Here, sets of observations with similar temperatures (previously determined individually) have been fitted simultaneously to reduce the uncertainties. Again, the corresponding data of the 2013 outburst (\hndk) are plotted in grey. All fits assume an absorbed blackbody parametrisation with the \nh fixed to the 2013 overall best fit of $1.4$ \hcm{21} (\hndk). A simultaneous modelling of all 2014 detections found best-fit parameters consistent with the 2013 results: $kT = (103^{+6}_{-5})$ eV and \nh = ($1.4\pm0.3$) \hcm{21}.

Figure \ref{fig:lc}b indicates that the effective temperature rose and fell together with the count rate. The peak blackbody temperature likely exceeded 110~eV, which no other nova in the \m31 sample of \md has yet come close to. The next highest, well determined $kT$ was $\sim80$~eV for M31N~2007-12b \citep[see also][]{2011A&A...531A..22P}. Because of the strong similarities between the 2013 and 2014 outbursts, we discuss the spectra of both campaigns together in Sect.\,\ref{sec:disc_spec}.

\section{Discussion}
\label{sec:discuss}
%

\subsection{Comparison to the 2013 outburst}
\label{sec:disc_2013}
The 2013 and 2014 \swift XRT light curves of \nova are strikingly similar (see Fig.\,\ref{fig:lc}a). The respective turn-on and turn-off times are consistent: $\eton = (6 \pm 1)$~d (2013) vs $\eton = (5.9 \pm 0.5)$~d (2014) and $\etoff = (19 \pm 1)$~d vs $\etoff = (18.4 \pm 0.5)$~d. The accurately measured \ton confirms our interpretation in \hnd of the SSS appearing around the time of the first 2013 observation. The 2014 time scale measurements presented in this work should be considered as having higher accuracy and precision.

Overall, the two X-ray light curves are consistent. While the 2013 light curve appears more variable, in contrast to the smoother plateau in 2014, this impression is caused by the longer exposure times of the 2014 observations. Both light curves show similar variability on short time scales. We discuss this in detail in Sect.\,\ref{sec:disc_var} and Fig.\,\ref{fig:xray_var}. The same figure shows that the 2014 count rates seem to be slightly higher than in 2013, albeit not by a statistically significant amount. Note, that also the optical light curves of the known outbursts are very similar (see \optk).

The X-ray spectral temperatures for both eruptions show a consistent evolution, too (see Fig.\,\ref{fig:lc}b). The longer exposure times of the 2014 monitoring led to higher numbers of photons and smaller statistical uncertainties on the estimated effective blackbody temperatures. In \hndk, we could not be certain of a spectral evolution. Here, the smaller errors of the 2014 measurements and the similarities to the 2013 results allow us to study the changes in effective temperature with considerably higher confidence by performing a combined analysis of all X-ray spectra from both years.

\subsection{Combined X-ray spectral evolution}
\label{sec:disc_spec}

\begin{figure}[t!]
  \resizebox{\hsize}{!}{\includegraphics[angle=0]{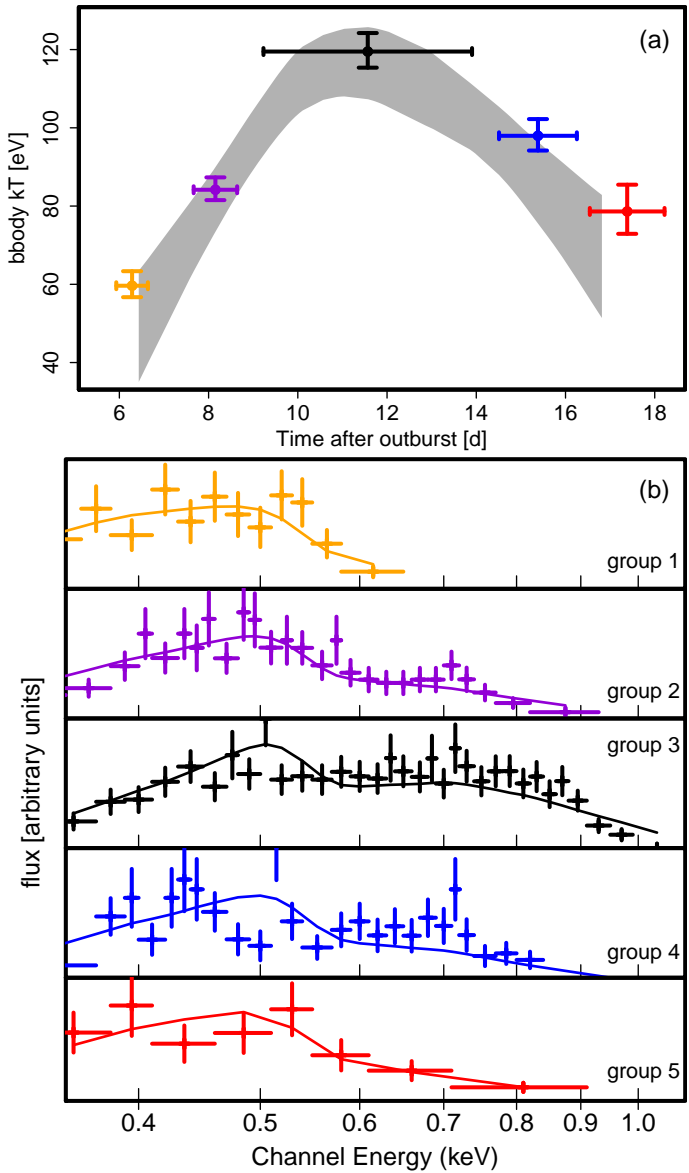}}
  \caption{\textit{Panel a}: The effective blackbody temperature of \nova depending on the time after outburst. Based on spectra from the 2014 and 2013 outburst (the latter described in \hndk). Sets of spectra with similar temperature (compare Fig.\,\ref{fig:lc}b) have been fitted simultaneously. Black/coloured data points show the best fit $kT$ and corresponding uncertainty. The error bars in time after outburst extend from the first to the last observation of each group. The grey region shows the 95\% confidence prediction interval derived from smoothing temperature fits based on individual snapshots (see Sect.\,\ref{sec:disc_spec} for details). \textit{Panel b}: Binned combined XRT spectra in arbitrary flux units with blackbody fits (solid lines). The colours correspond to the outburst stages in panel a.}
  \label{fig:temp}
\end{figure}

First, let the reader be reminded that it is widely acknowledged that blackbody parametrisations of X-ray spectra of novae do not produce physical results. High-resolution X-ray spectra of Galactic novae during the SSS phase have revealed a wealth of absorption or emission lines superimposed on the continuum \citep[e.g.][]{2012BASI...40..353N,2013A&A...559A..50N}. Non-LTE atmosphere models, if possible including atmosphere expansion, have been suggested for the accurate modelling of the SSS state of novae \citep[e.g.][]{2005A&A...431..321P,2010ApJ...717..363R,2012ApJ...756...43V}.

However, this work is based on low-count, low-resolution CCD spectra obviously insufficient to impose any credible constraints to atmosphere models. Our analysis does not aim to provide accurate absolute temperatures, but rather to measure the relative temperature evolution during and between the outbursts. We are employing the blackbody model as a simple and robust parametrisation of the spectral shape. In a similar way, blackbody fits were used by \md to model the population properties of \m31 novae, to which we compare \nova in Sect.\,\ref{sec:disc_pop} below. In the present work, relative changes in the spectrum are easier to study, because we are comparing data from the same instrument of the same nova.

We combined the sets of spectra used to produce Fig.\,\ref{fig:lc}b into the five groups that most readers might intuitively construct from this figure. These groups correspond to the stages of the rise and decline of the light curve (two groups each; also compare Fig.\,\ref{fig:lc}a) as well as the light curve plateau. The latter is represented by one group comprising two (black) sets of 2014 spectra and three (grey) sets of 2013 spectra (compare Fig.\,\ref{fig:lc}b). Groups of spectra were fitted simultaneously, again assuming a fixed \nh = $1.4$ \hcm{21}.

The resulting effective temperature evolution is shown by the black/coloured data points in Fig.\,\ref{fig:temp}. We clearly see an increase in effective temperature during the first four days of the SSS phase, from $60 \pm 4$~eV (orange; group 1) over $(84 \pm 3)$~eV (purple; group 2) to the maximum of $120 \pm 5$~eV (black; group 3), while the count rate was still increasing as well. The maximum effective temperature was reached during the light curve plateau. Subsequently, the blackbody temperatures decreased to $(98 \pm 4)$~eV (blue; group 4) and finally $79 \pm 7$~eV (red; group 5) just before the count rate dropped below our detection limit.

We also performed a blackbody parametrisation of the spectra of individual snapshots. This included only observations with more than 200~s exposure time for which a stable fit could be obtained. In total, we used 63 individual spectra (28 from 2014 and 35 from 2013). These individual fits were subjected to a smoothing using local polynomial regression filtering with the LOESS method \citep{loess_Cleveland92}. The fitting used least squares weighted by the inverse of the variances. In Fig.\,\ref{fig:temp} we also show the 95\% confidence prediction intervals of this smoothing fit as a grey band. For reasons of clarity the individual snapshot temperatures are not shown.

The grouped fits and smoothed regression are consistent in showing a significant rise in effective temperature. This is consistent with theoretical models predicting a hardening of the photospheric emission due to the gradual emergence of hotter atmosphere layers \citep[e.g.][]{2005A&A...439.1061S,2006ApJS..167...59H,2013ApJ...777..136W}. The high blackbody temperature maximum of $120 \pm 5$~eV is unprecedented in the \m31 nova sample (see \md and Sect.\,\ref{sec:disc_pop}). This indicates an extremely massive WD, following the correlation between effective temperatures and WD masses in the models of \citet{2005A&A...439.1061S} and \citet{2013ApJ...777..136W} that suggest that more massive WDs on average exhibit higher temperatures. A high WD mass would be consistent with the fast SSS time scales (see Sect.\,\ref{sec:disc_var}) and short outburst recurrence time \citep{2014ApJ...793..136K}.

During the decline of the SSS flux there is clear evidence for a decreasing effective temperature (compare Figs.\,\ref{fig:lc} and \ref{fig:temp}). While there were already indications for this behaviour in the 2013 data we presented in \hndk, adding the new 2014 measurements provides a considerably higher level of certainty. The cooling is consistent with predictions by theoretical models \citep{2005A&A...439.1061S,2013ApJ...777..136W}.

The drop in luminosity from the plateau to the last detections is consistent with a $T^4$ dependency on the photospheric temperature drop (at constant radius) between the third and the last fit in Fig.\,\ref{fig:temp} (both a factor of $\sim 5$). In contrast, during the rise of the SSS light curve the average count rate increased considerably less than the blackbody temperature (factor $\sim3$ vs factor $\sim16$ for $T^4$). This behaviour is predicted by theoretical models that assume a spectral hardening caused by a receding photosphere \citep[e.g.][]{2005A&A...439.1061S}. In this scenario, the decreasing radius would reduce the rise in flux due to the increasing effective temperature.

In Fig.\,\ref{fig:temp}b we show the merged, binned XRT spectra of the five groups in Fig.\,\ref{fig:temp}a together with a blackbody fit. This provides a visualisation of the spectral evolution. Note, that the temperature fits for Fig.\,\ref{fig:temp}a were performed on the spectra of individual observations using Poisson statistics according to \citet{1979ApJ...228..939C}. The binned spectra are broadly consistent with the temperature evolution suggested in Fig.\,\ref{fig:temp}a. However, they also suggest additional spectral features beyond a simple blackbody continuum when comparing the fit to the data, e.g. at around 0.7~keV in groups 2 and 4 or to the flat top spectrum of group 3.

Any more detailed interpretation is not possible given the quality of our spectra. High-resolution X-ray spectra of Galactic novae \citep[see e.g.][]{2012BASI...40..353N,2013A&A...559A..50N} show that their SSS state can show a plethora of emission and absorption features \citep[see e.g. the RN U~Sco]{2012ApJ...745...43N}. Similarly, some parts of the X-ray spectrum that here we attribute to the blackbody continuum could be due to overlapping, unresolved emission or absorption features.

\subsection{Combined X-ray light curve and variability}
\label{sec:disc_var}
In \hnd we found significant X-ray variability during the SSS phase of \novak. In contrast, the 2014 light curve appears smoother (compare the grey and black data in Fig.\,\ref{fig:lc}a). However, the 2014 exposures were typically longer than in 2013 (see Table\,\ref{tab:obs} and \hndk) and would therefore produce an average count rate that could hide short term variability. To study this variability we estimated separately the count rates for all the individual \swift XRT snapshots of the 2014 monitoring. The resulting count rates are given in Table\,\ref{tab:obs_split}. We also revisited the 2013 data, as described in \hndk, and applied the same procedure to extract the count rates given in Table\,\ref{tab:obs_split13}.

%
\begin{figure}[t!]
  \resizebox{\hsize}{!}{\includegraphics[angle=0]{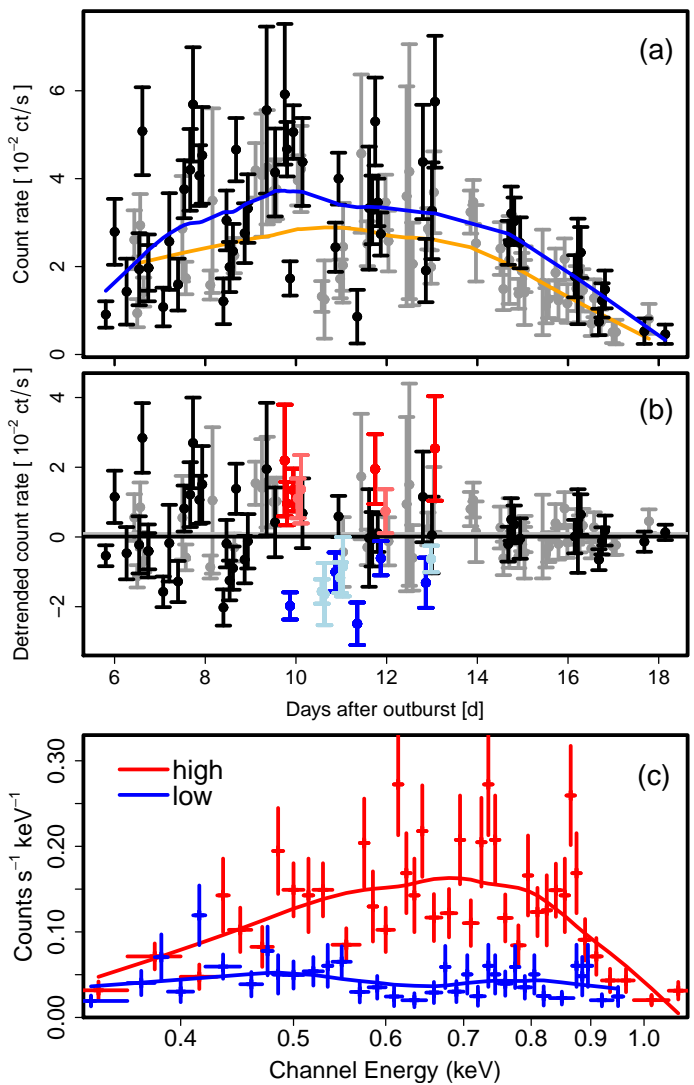}}
  \caption{\textit{Panel a}: The short-term X-ray light curve of \nova based on the individual XRT snapshots. Data points with error bars show the XRT count rates and corresponding errors (see Table\,\ref{tab:obs_split}) for the 2014 (black) and 2013 (grey) outbursts. Note, that the count rate axis uses a linear scale in contrast to the logarithmic scale in Fig.\,\ref{fig:lc}. Solid lines represent smoothed fits, based on local regression, on the 2014 (blue) and 2013 (orange) data. \textit{Panel b}: The light curves from panel a have been de-trended by subtracting the smoothed fits from the respective data. The red (and light red) data points mark the count rates that are at least $1\sigma$ above the smoothed fit for the 2014 (2013) data during the temperature maximum. The blue (and light blue) data are at least $1\sigma$ below the average 2014 (2013) count rate for the same time range. \textit{Panel c}: Binned XRT spectra for all the high (red colours) and low luminosity (blue colours) snapshots of the 2014 and 2013 monitoring that are indicated in panel b with corresponding colours.}
  \label{fig:xray_var}
\end{figure}

In Fig.\,\ref{fig:xray_var}a we show the short term X-ray light curves of \nova in 2014 (black) and 2013 (grey). This plot clearly reveals that both light curves show a similar amount of variability on time scales of hours. Smoothed fits, based on the LOESS local regression and shown as solid lines in Fig.\,\ref{fig:xray_var}a, suggest that the 2014 outburst (blue) was on average slightly brighter than the 2013 outburst (orange). The difference in luminosity is not statistically significant, but it is consistent with indications that the effective temperature of the 2014 outburst in the later stages might have been systematically higher than in 2013 (see Fig.\,\ref{fig:lc}b). Observations of future outbursts will allow us to put useful constraints on the variations in temperature and luminosity between outbursts.

The X-ray variability of \nova can best be seen in Fig.\,\ref{fig:xray_var}b, where we show the de-trended light curves of the 2014 (black, red, blue) and 2013 (grey, light red, light blue) SSS phases. The de-trending assumes the smoothed average shown in Fig.\,\ref{fig:xray_var}a, which has been subtracted. In Fig.\,\ref{fig:xray_var}b two things are immediately obvious: Both light curves show (a) a similar degree of variability which (b) is reduced considerably after about day 13. This visual impression is confirmed by statistical F-tests: the variances in XRT count rate of the 2013 and 2014 de-trended light curves before day 13 are not significantly different (1.1 vs 1.7). The same is true for the variances after day 14 (0.12 vs 0.13). However, the difference between the high- and low-variability parts of both light curves are highly significant beyond the $3\sigma$ level with p-values of 2.1\tpower{-6} (2013) and 1.8\tpower{-5} (2014).

The significant drop in variability after day 13 could indicate the end of an early variability phase which has been observed in a number of Galactic novae \citep[e.g. KT~Eri, RS~Oph;][]{2010ATel.2392....1B,2011ApJ...727..124O}. We performed a Lomb-Scargle analysis \citep{1976Ap&SS..39..447L,1982ApJ...263..835S} on the two de-trended high-cadence light curves in Fig.\,\ref{fig:xray_var}b before day 14. We did not detect any periodic signal in either light curve on the 95\% confidence level.

Additionally, we searched for any characteristic variability time scales with an approach similar to the structure function described by \citet{1985ApJ...296...46S}. For any two pairs of observations we recorded the time lag and the difference in count rate. We then applied a local regression smoother to estimate the average variability on different time scales. We found the resulting smoothed average to be flat over time; i.e. all time scales (from seven minutes to seven days) appear to show a similar degree of variability.

To shed further light on the origins of the high-variability phase we compared the XRT spectra of those measurements significantly above and below the average (smoothed) count rate. We further restricted this comparison to snapshots taken during the temperature maximum (the black data point in Fig.\,\ref{fig:temp}a), to avoid a contamination by the overall temperature evolution. The selected high-flux data points are marked in red (2014) or light red (2013) in Fig.\,\ref{fig:xray_var}b, while the low-flux measurements are coloured blue (2014) or light blue (2013). The merged and binned spectra are shown with corresponding colours in Fig.\,\ref{fig:xray_var}c.

The largest difference between the two spectra seems to occur in the energy range of $0.6 - 0.8$~keV, while particularly at lower energies the spectra become more similar. This could indicate that the variability is not caused by absorption in neutral material, which should affect softer X-rays more. Absorption by completely opaque material appears unlikely as well, because this would not change the shape of the spectrum. Changes in photospheric temperature are more likely to show effects as in Fig.\,\ref{fig:xray_var}c (see Ness et al. 2015, in prep.).

The high-flux data points (red) in Fig.\,\ref{fig:xray_var}b have count rates that are on average a factor of $2.5$ higher than for the low-flux (blue) data points. This would correspond to a factor of $1.3$ higher temperatures if caused by a $T^4$ dependency (e.g. 105~kT to 133~kT). Other possible explanations are an emission line dominated spectrum with variable line strengths (in the $0.6-0.8$~keV range most likely oxygen lines) or a varying degree of ionisation of O\,{\sc i} \citep[as suggested for RS~Oph in][]{2015arXiv150104791N}.

\subsection{Multiwavelength evolution}
\label{sec:disc_multi}
\begin{figure}[t!]
  \resizebox{\hsize}{!}{\includegraphics[angle=0]{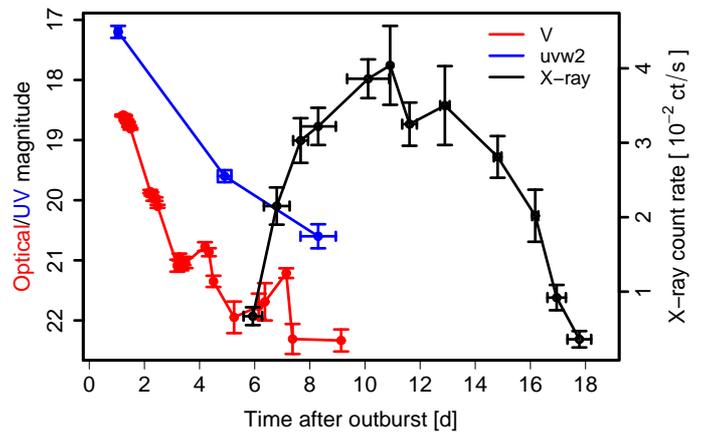}}
  \caption{Optical (red, $V$ filter), UV (blue, $uvw2$ filter) and X-ray light curves of \nova during the 2014 outburst. The optical/UV magnitudes were taken from \opt (see also Table\,\ref{tab:obs}).}
  \label{fig:multi_lc}
\end{figure}

Theoretical nova models predict a gradual hardening of the source spectrum that ultimately leads to the emergence of the SSS \citep[e.g.][]{2006ApJS..167...59H}. In Fig.\,\ref{fig:multi_lc} we visualise the overall multiwavelength evolution of \nova using three energy bands: the optical ($V$ band), the UV ($uvw2$ filter), and the soft X-ray regime. The optical and UV measurements were taken from \optk, where more detailed light curves of the 2014 outburst in various optical/UV filters are presented and discussed.

In Fig.\,\ref{fig:multi_lc} we see the $V$ band and UV magnitudes declining in a similar way. There is no significant shift in emission from the optical to the UV, which is bright already very early in the outburst. Together with the faint optical peak magnitude this behaviour is consistent with a weak outburst in which the expanding pseudo photosphere exhibits a high effective temperature and never reaches a red giant size (see \hnd and \optk).

The $V$ band light curve shows two distinctive re-brightenings, the latter of which coincides with the rising SSS emission. \citet{2008ASPC..401..206H} suggested that the observed plateaus in the optical light curves of Galactic novae like RS~Oph are caused by the reprocessing of X-ray emission by the accretion disk. Here, reprocessed SSS emission might have slowed down, and even briefly reversed, the optical light curve decay. We cannot exclude that an optical plateau did occur below the sensitivity limit of the observations. See \opt for a detailed discussion of the optical light curve.

%

\subsection{Updated population picture, ejected hydrogen mass, \& RN time scales}
\label{sec:disc_pop}
\begin{figure}[t!]
  \resizebox{\hsize}{!}{\includegraphics[angle=0]{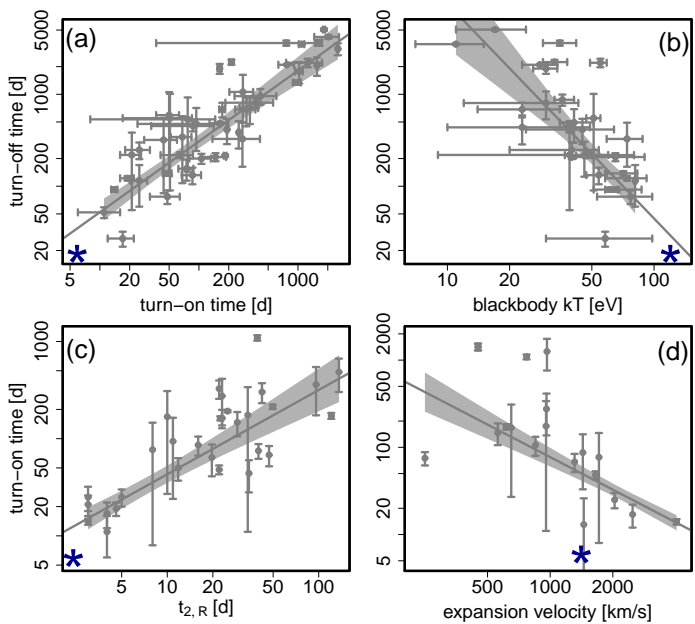}}
  \caption{Double-logarithmic plots of the nova parameter correlations from \mdk. The solid grey lines indicate a robust power-law fit with corresponding 95\% confidence regions in light grey. The correlations displayed are (a) \ton vs \toffk, (b) black-body $kT$ vs \toffk, (c) $R$-band optical decay time $t_{2,R}$ vs \tonk, and (d) expansion velocity vs \tonk. All time scales are in units of days after outburst. Overplotted as blue asterisks are the parameters of \nova as derived in Sect.\,\ref{sec:results} or by \optk. The error bars on these values are smaller than the size of the symbols.}
  \label{fig:corr}
\end{figure}

In Fig.\,\ref{fig:corr} we provide an update on the connection of \nova with the big picture of \m31 novae as it was first discussed in \hndk. The data for the \m31 nova sample (grey) is based on \mdk, who discussed the four correlations and their implications in detail. Here, we assume the 2014 measurements for the optical \ttwo decline rate and the expansion velocity as presented by \optk. We further use the updated SSS time scales, \ton and \toffk, as derived in Sect.\,\ref{sec:res_lc}, and assume the peak blackbody temperature ($kT = 120 \pm 5$~eV) estimated in Sect.\,\ref{sec:disc_spec}.

All parameters are in good agreement with the average multiwavelength behaviour of \m31 novae, except for the expansion velocity. A low expansion velocity (and a faint optical maximum; see \optk) could indicate a weak outburst on an extremely high mass WD (see \hndk). We note that this finding is at odds with a recent study on Galactic novae proposing high expansion velocities as one among several promising indicators for an underlying RN \citep{2014ApJ...788..164P}. \nova shows clearly that low expansion velocities do not rule out an RN. This is in agreement with theoretical models predicting lower expansion velocities for very high accretion rates \citep[plus short recurrence times in case of high-mass WDs, see][]{2005ApJ...623..398Y}.

The remaining three parameter connections in Fig.\,\ref{fig:corr}, while consistently below the \m31 average, are still in good agreement with the overall population fits and therefore extend these relations to faster time scales and higher temperatures. We note, that \nova is not the nova with the fastest SSS \ton any more. This title was taken by the Galactic RN V745~Sco in February 2014 \citep[SSS turn on 3-4 d, see][]{2014ATel.5870....1P}.

%

Based on the SSS $\eton =  (5.9 \pm 0.5)$~d we estimated the hydrogen mass ejected during the outburst. This assumes that the central SSS becomes visible once the ejecta turn optically thin to soft X-rays. Following \citet{2002A&A...390..155D}, we assumed the ejecta to expand as a spherical symmetric shell with a thickness/filling factor of 0.2 \citep[see also][for a detailed description of the method]{2010A&A...523A..89H}. We used the average H$\alpha$ FWHM = 2500~km~s$^{-1}$ (see \optk) to derive the corresponding expansion velocity as the standard deviation of the Gaussian line. We found an ejected hydrogen mass of $\emej = (2.6 \pm 0.4)$~\tpower{-8} \msun. This is consistent with the overall ejecta mass estimated from the optical data in  \optk.

\begin{figure}
 \resizebox{\hsize}{!}{\includegraphics{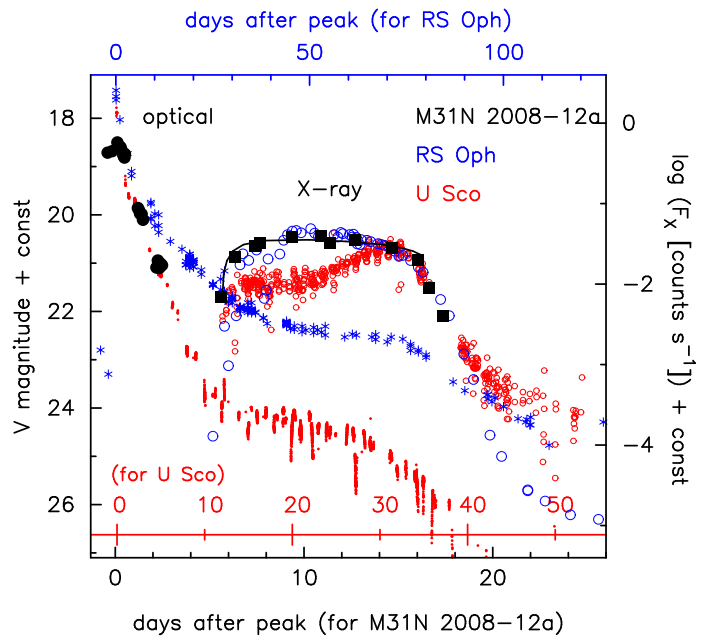}} 
 \caption{Comparison of the light curves of \nova (black) with the Galactic RNe RS~Oph (blue) and U~Sco (red) in normalised time scales. Filled black circles, for \novak, indicate $V$ magnitudes as presented in \opt and filled black squares show the X-ray count rates (compare Fig.\,\ref{fig:lc} and Table\,\ref{tab:obs}). Blue asterisks, for RS~Oph, give the optical ($V$, visual and $y$) magnitudes taken from \citet{2006ApJ...651L.141H} and blue open circles indicate the X-ray count rates (0.3 - 0.55 keV) taken from \citet{2007ApJ...659L.153H}. Red dots, for U~Sco, show the optical magnitudes taken from the AAVSO archive while small open red circles give the X-ray count rate (0.2--1.0 keV) deduced from the \swift archive. The black solid line indicates a theoretical light curve for the SSS blackbody flux from a $1.377~M_\odot$ WD with a chemical composition of $X=0.6$, $Y=0.38$, and $Z=0.02$ between $t_{\rm on}^{\rm theory}$ and $t_{\rm off}^{\rm theory}$.}
 \label{fig:m31_gal}
\end{figure}

Furthermore, in Fig.\,\ref{fig:m31_gal} we compare the optical and X-ray light curves of \nova to those of the two prominent Galactic RNe RS~Oph (2006 outburst) and U~Sco (2010 outburst). Although the three RNe had different outburst durations of 18~d for \novak, 35~d for U~Sco, and 83~d for RS~Oph, we show that their overall light curve evolutions looks remarkably similar after a linear scaling. These similarities are particularly striking for the early optical decline, the SSS turn-on and turn-off times, and the X-ray cooling times after the SSS turn off.

The scaling behaviour of RNe is in contrast to CNe, where the time-scalability of the optical decay can be described by a universal decline law which predicts a different linear proportionality, i.e. a different slope, between $t_2$ vs. \ton and $t_2$ vs. \toff \citep[see Figures 12 and 13 in][]{2010ApJ...709..680H}. This is because the early optical decline depends directly on the WD mass but barely on the hydrogen content $X$ in the envelope which is fuelling the SSS phase. This difference in the proportionality is qualitatively consistent with the observed behaviour of the population trends for \m31 novae (see \md and Fig.\,\ref{fig:corr}). Thus, the scalability of these RNe might be a clue to understanding the physics of novae with very short recurrence periods. We note that \citet{2014arXiv1407.7076I} reported that also the Galactic RN T~Pyx in its 2011 outburst showed SSS time scales consistent with the \m31 trends. A systematic comparison between novae in \m31 and the Galaxy is in preparation.

In Fig.\,\ref{fig:m31_gal} we also included a theoretical SSS light curve calculated following \citet{1994ApJ...437..802K,1999PASJ...51..525K,2006ApJS..167...59H} for a $1.377~M_\odot$ WD with an envelope chemical composition of $X=0.6$, $Y=0.38$, and $Z=0.02$. The X-ray light curve starts when the optically thick wind stops (theoretical SSS turn-on time $t_{\rm on}^{\rm theory}$) and ends at the epoch when hydrogen shell-burning extinguishes (theoretical SSS turn-off time $t_{\rm off}^{\rm theory}$). To fit the count rates of \novak, we freely shifted the theoretical light curve vertically. We conclude that the SSS phase of \nova is consistent with our theoretical model of a $1.377~M_\odot$ WD representing the upper mass end of mass-accreting and non-rotating WDs.

%

\section{Summary \& Conclusions}
\label{sec:summary}
%
We carried out high-cadence \swift XRT monitoring of the 2014 outburst of the remarkable recurrent nova \novak. These are our main findings:

\begin{enumerate}

  \item \nova was detected as a bright SSS with well constrained and very fast time scales of $\eton =  (5.9 \pm 0.5)$~d and $\etoff = (18.4 \pm 0.5)$~d.\\

  \item The X-ray light curve and spectral evolution of the 2014 outburst are very similar to the 2013 outburst presented in \hndk.\\

  \item The combined 2014 and 2013 spectral evolution suggests a fast rise in temperature followed by an equally rapid decline. Based on a blackbody parametrisation we estimated a maximum effective temperature of $kT = 120 \pm 5$~eV, significantly above any other \m31 nova.\\

  \item The short-term X-ray light curves of 2014 and 2013 display strong, aperiodic variability that decreased significantly around day 14 after outburst.\\

  \item The X-ray parameters of \nova fit well into the big picture of the \m31 sample presented by \mdk. The connection between optical and X-ray time scales is consistent with the average \m31 nova behaviour as well. Only the expansion velocity, as measured by \optk, is lower than expected, which might be explained by a weak outburst on a high-mass WD with a high accretion rate.\\
  
  \item The SSS phase of \nova can be modelled assuming a WD close to the Chandrasekhar mass. The optical and X-ray light curve can be scaled to show similar time scales as the light curves of the Galactic RNe RS~Oph and U~Sco.

\end{enumerate}

\nova offers a unique laboratory for detailed studies of repeated nova outbursts. For the first time, the unprecedented one-year recurrence time makes it feasible to follow a single nova over a large number of outbursts. We are confident that \nova will show further outbursts in the foreseeable future and we predict the next outburst for the autumn of 2015 which we aim to study in great detail.

\begin{acknowledgements}
We are grateful to the \swift Team for the rapid scheduling of the ToO observations, in particular N. Gehrels, the duty scientists, as well as the science planners. M. Henze acknowledges support from an ESA fellowship. A.W.S. acknowledges support from NSF grant AST1009566. G.S. acknowledges the support of the Spanish Ministry of Economy and Competitivity (MINECO) under the grant AYA2011-23102. M.K. and I.H. acknowledge support from the Grants-in-Aid for Scientific Research (22540254, 24540227) of the Japan Society for the Promotion of Science. M. Hernanz acknowledges the support of the Spanish Ministry of Economy and Competitivity (MINECO) under the grant ESP2013-41268-R.

\end{acknowledgements}

\bibliographystyle{aa}

%
\begin{table*}[ht]
\caption{\swift observations of nova \nova following the 2014 outburst.}
\label{tab:obs}
\begin{center}
\begin{tabular}{rrrrrrrrrrrr}\hline\hline \noalign{\smallskip}
ObsID & Exp$^a$ & Date$^b$ & MJD$^b$ & $\Delta t^c$ & \multicolumn{3}{c}{UV$^d$ [mag]} & Rate &  L$_{0.2-1.0}$ $^e$\\
& [ks] & [UT] & [d] & [d] & uvm2 & uvw1 & uvw2 & [\power{-2} ct s$^{-1}$] & \power{38} erg s$^{-1}$]\\ \hline \noalign{\smallskip}
00032613042 & 7.9 & 2014-10-03.63 & 56933.63 & 0.94 & - & - & $17.2\pm0.1$ & $<0.2$ & $<0.2$ \\
00032613043 & 7.9 & 2014-10-04.44 & 56934.44 & 1.75 & $18.3\pm0.1$ & - & - & $<0.2$ & $<0.1$ \\
00032613045 & 3.9 & 2014-10-05.16 & 56935.16 & 2.47 & - & $19.4\pm0.1$ & - & $<0.4$ & $<0.3$ \\
00032613046 & 1.8 & 2014-10-06.02 & 56936.02 & 3.33 & - & - & - & $<0.6$ & $<0.4$ \\
00032613047 & 7.3 & 2014-10-07.35 & 56937.35 & 4.66 & - & - & $19.6\pm0.1$ & $<0.2$ & $<0.2$ \\
00032613048 & 6.9 & 2014-10-08.29 & 56938.29 & 5.60 & $20.4\pm0.2$ & - & - & $0.7\pm0.1$ & $0.5\pm0.1$ \\
00032613049 & 4.4 & 2014-10-09.02 & 56939.02 & 6.33 & - & $20.0\pm0.2$ & - & $2.2\pm0.2$ & $1.6\pm0.2$ \\
00032613051 & 4.4 & 2014-10-10.09 & 56940.09 & 7.40 & - & - & - & $3.0\pm0.3$ & $2.3\pm0.2$ \\
00032613050 & 6.5 & 2014-10-10.35 & 56940.35 & 7.66 & - & - & $20.6\pm0.2$ & $3.2\pm0.2$ & $2.4\pm0.2$ \\
00032613052 & 7.7 & 2014-10-12.04 & 56942.04 & 9.35 & $21.1\pm0.4$ & $>20.1$ & - & $3.9\pm0.3$ & $2.9\pm0.2$ \\
00032613054 & 1.8 & 2014-10-13.56 & 56943.57 & 10.88 & - & $>20.4$ & - & $4.0\pm0.5$ & $3.0\pm0.4$ \\
00032613055 & 5.3 & 2014-10-14.04 & 56944.04 & 11.35 & - & - & - & $3.2\pm0.3$ & $2.4\pm0.2$ \\
00032613056 & 1.6 & 2014-10-15.42 & 56945.42 & 12.73 & - & - & $>20.6$ & $3.5\pm0.5$ & $2.6\pm0.4$ \\
00032613057 & 4.8 & 2014-10-17.36 & 56947.36 & 14.67 & - & $>20.9$ & - & $2.8\pm0.3$ & $2.1\pm0.2$ \\
00032613058 & 2.2 & 2014-10-18.75 & 56948.75 & 16.06 & - & - & - & $2.0\pm0.4$ & $1.5\pm0.3$ \\
00032613059 & 4.1 & 2014-10-19.31 & 56949.31 & 16.62 & - & - & $>21.2$ & $0.9\pm0.2$ & $0.7\pm0.1$ \\
00032613060 & 4.7 & 2014-10-20.03 & 56950.04 & 17.35 & $21.0\pm0.4$ & - & - & $0.4\pm0.1$ & $0.3\pm0.1$ \\
00032613061 & 5.8 & 2014-10-21.42 & 56951.42 & 18.73 & - & $>21.1$ & - & $<0.3$ & $<0.2$ \\
00032613062 & 6.1 & 2014-10-22.03 & 56952.03 & 19.34 & - & - & - & $<0.2$ & $<0.2$ \\
\hline
\end{tabular}
\end{center}
\noindent
Notes:\hspace{0.1cm} $^a $: Dead-time corrected exposure time; $^b $: Start date of the observation; $^c $: Time in days after the outburst of nova \nova in the optical on 2014-10-02.69 UT (MJD 56932.69; see Sect.\,\ref{sec:res_lc}); $^d $: \swift UVOT filters were UVM2 (166-268nm), UVW1 (181-321nm), and UVW2 (112-264nm); $^e $:X-ray luminosities (unabsorbed, black-body fit, 0.2 - 10.0 keV) and upper limits were estimated according to Sect.\,\ref{sec:results}.\\
\end{table*}

\clearpage

%
\begin{table*}[ht]
\caption{Individual \swift snapshots of the observations in Table\,\ref{tab:obs_split}.}
\label{tab:obs_split}
\begin{center}
\begin{tabular}{rrrrrrrrrrrr}\hline\hline \noalign{\smallskip}
ObsID\_part & Exp$^a$ & Date$^b$ & MJD$^b$ & $\Delta t^c$ & \multicolumn{3}{c}{UV$^d$ [mag]} & Rate &  L$_{0.2-1.0}$ $^e$\\
& [ks] & [UT] & [d] & [d] & uvm2 & uvw1 & uvw2 & [\power{-2} ct s$^{-1}$] & \power{38} erg s$^{-1}$]\\ \hline \noalign{\smallskip}
00032613042\_1 & 1.59 & 2014-10-03.63 & 56933.63 & 0.94 & - & - & $17.0\pm0.1$ & $<0.4$ & $<0.3$ \\
00032613042\_2 & 2.46 & 2014-10-03.69 & 56933.69 & 1.00 & - & - & $17.0\pm0.1$ & $<0.5$ & $<0.4$ \\
00032613042\_3 & 2.46 & 2014-10-03.76 & 56933.75 & 1.06 & - & - & $17.3\pm0.1$ & $<0.5$ & $<0.4$ \\
00032613042\_4 & 1.45 & 2014-10-03.82 & 56933.82 & 1.13 & - & - & $17.2\pm0.1$ & $<1.1$ & $<0.8$ \\
00032613043\_1 & 0.74 & 2014-10-04.44 & 56934.44 & 1.75 & $18.1\pm0.1$ & - & - & $<1.0$ & $<0.7$ \\
00032613043\_2 & 1.75 & 2014-10-04.50 & 56934.50 & 1.81 & $18.1\pm0.1$ & - & - & $<0.5$ & $<0.4$ \\
00032613043\_3 & 1.70 & 2014-10-04.56 & 56934.56 & 1.87 & $18.2\pm0.1$ & - & - & $<0.7$ & $<0.5$ \\
00032613043\_4 & 2.53 & 2014-10-04.62 & 56934.62 & 1.93 & $18.4\pm0.1$ & - & - & $<0.3$ & $<0.3$ \\
00032613043\_5 & 0.90 & 2014-10-04.69 & 56934.69 & 2.00 & $18.3\pm0.1$ & - & - & $<0.9$ & $<0.7$ \\
00032613043\_6 & 0.33 & 2014-10-04.71 & 56934.71 & 2.02 & $18.6\pm0.2$ & - & - & $<2.2$ & $<1.7$ \\
00032613045\_1 & 0.87 & 2014-10-05.16 & 56935.16 & 2.47 & - & $19.1\pm0.2$ & - & $<1.4$ & $<1.0$ \\
00032613045\_2 & 0.29 & 2014-10-05.31 & 56935.30 & 2.61 & - & $19.0\pm0.3$ & - & $<4.4$ & $<3.3$ \\
00032613045\_3 & 1.01 & 2014-10-05.62 & 56935.62 & 2.93 & - & $19.4\pm0.2$ & - & $<0.9$ & $<0.7$ \\
00032613045\_4 & 1.10 & 2014-10-05.69 & 56935.69 & 3.00 & - & $19.5\pm0.3$ & - & $<1.5$ & $<1.1$ \\
00032613045\_5 & 0.60 & 2014-10-05.82 & 56935.82 & 3.13 & - & $>19.6$ & - & $<1.4$ & $<1.0$ \\
00032613046\_1 & 0.02 & 2014-10-06.02 & 56936.02 & 3.33 & - & - & - & $<38.4$ & $<28.8$ \\
00032613046\_2 & 0.14 & 2014-10-06.50 & 56936.50 & 3.81 & - & - & - & $<5.4$ & $<4.1$ \\
00032613046\_3 & 1.23 & 2014-10-06.89 & 56936.89 & 4.20 & - & - & - & $<0.8$ & $<0.6$ \\
00032613046\_4 & 0.38 & 2014-10-06.95 & 56936.95 & 4.26 & - & - & - & $<2.3$ & $<1.7$ \\
00032613047\_1 & 0.13 & 2014-10-07.35 & 56937.36 & 4.67 & - & - & $>18.8$ & $<6.5$ & $<4.8$ \\
00032613047\_2 & 0.82 & 2014-10-07.36 & 56937.36 & 4.67 & - & - & $>20.2$ & $<1.0$ & $<0.7$ \\
00032613047\_3 & 1.72 & 2014-10-07.42 & 56937.42 & 4.73 & - & - & $19.6\pm0.2$ & $<0.6$ & $<0.5$ \\
00032613047\_4 & 1.71 & 2014-10-07.49 & 56937.49 & 4.80 & - & - & $19.0\pm0.1$ & $<0.4$ & $<0.3$ \\
00032613047\_5 & 1.05 & 2014-10-07.57 & 56937.57 & 4.88 & - & - & $19.6\pm0.2$ & $<0.9$ & $<0.7$ \\
00032613047\_6 & 1.55 & 2014-10-07.63 & 56937.62 & 4.93 & - & - & $19.9\pm0.2$ & $<0.7$ & $<0.5$ \\
00032613047\_7 & 0.37 & 2014-10-07.84 & 56937.84 & 5.15 & - & - & $19.4\pm0.3$ & $<3.5$ & $<2.6$ \\
00032613048\_1 & 0.88 & 2014-10-08.29 & 56938.29 & 5.60 & $19.7\pm0.3$ & - & - & $<1.6$ & $<1.2$ \\
00032613048\_2 & 1.00 & 2014-10-08.35 & 56938.36 & 5.67 & $>20.0$ & - & - & $<1.3$ & $<1.0$ \\
00032613048\_3 & 0.96 & 2014-10-08.42 & 56938.43 & 5.74 & $19.8\pm0.3$ & - & - & $<0.8$ & $<0.6$ \\
00032613048\_4 & 1.51 & 2014-10-08.49 & 56938.49 & 5.80 & $20.1\pm0.3$ & - & - & $0.9\pm0.3$ & $0.7\pm0.2$ \\
00032613048\_5 & 0.97 & 2014-10-08.55 & 56938.55 & 5.86 & $>20.0$ & - & - & $<1.6$ & $<1.2$ \\
00032613048\_6 & 0.60 & 2014-10-08.62 & 56938.62 & 5.93 & $>19.7$ & - & - & $<2.9$ & $<2.2$ \\
00032613048\_7 & 0.61 & 2014-10-08.69 & 56938.69 & 6.00 & $>19.7$ & - & - & $2.8\pm0.8$ & $2.1\pm0.6$ \\
00032613048\_8 & 0.38 & 2014-10-08.95 & 56938.95 & 6.26 & $>19.4$ & - & - & $1.4\pm0.8$ & $1.1\pm0.6$ \\
00032613049\_1 & 0.34 & 2014-10-09.02 & 56939.02 & 6.33 & - & $>19.4$ & - & $<5.7$ & $<4.3$ \\
00032613049\_2 & 0.39 & 2014-10-09.23 & 56939.23 & 6.54 & - & $>19.5$ & - & $1.9\pm0.8$ & $1.5\pm0.6$ \\
00032613049\_3 & 0.60 & 2014-10-09.30 & 56939.30 & 6.61 & - & $>19.7$ & - & $5.1\pm1.0$ & $3.8\pm0.8$ \\
00032613049\_4 & 0.43 & 2014-10-09.43 & 56939.43 & 6.74 & - & $20.1\pm0.3$ & - & $2.0\pm0.8$ & $1.5\pm0.6$ \\
00032613049\_5 & 0.93 & 2014-10-09.43 & 56939.43 & 6.74 & - & - & - & $2.0\pm0.5$ & $1.5\pm0.4$ \\
00032613049\_6 & 0.23 & 2014-10-09.55 & 56939.55 & 6.86 & - & $>19.0$ & - & $<6.2$ & $<4.6$ \\
00032613049\_7 & 0.44 & 2014-10-09.69 & 56939.69 & 7.00 & - & $>19.6$ & - & $<3.5$ & $<2.6$ \\
00032613049\_8 & 0.74 & 2014-10-09.76 & 56939.76 & 7.07 & - & $19.6\pm0.3$ & - & $1.1\pm0.4$ & $0.8\pm0.3$ \\
00032613049\_9 & 0.28 & 2014-10-09.90 & 56939.89 & 7.20 & - & $>18.6$ & - & $2.6\pm1.1$ & $1.9\pm0.8$ \\
00032613051\_1 & 0.63 & 2014-10-10.09 & 56940.09 & 7.40 & - & - & - & $1.6\pm0.6$ & $1.2\pm0.4$ \\
00032613051\_2 & 0.93 & 2014-10-10.15 & 56940.15 & 7.46 & - & - & - & $1.0\pm0.4$ & $0.7\pm0.3$ \\
00032613051\_3 & 1.25 & 2014-10-10.22 & 56940.22 & 7.53 & - & - & - & $3.8\pm0.7$ & $2.8\pm0.5$ \\
00032613051\_4 & 0.04 & 2014-10-10.29 & 56940.29 & 7.60 & - & - & - & $<21.0$ & $<15.8$ \\
00032613050\_1 & 0.65 & 2014-10-10.35 & 56940.36 & 7.67 & - & - & - & $4.2\pm0.9$ & $3.2\pm0.7$ \\
00032613050\_2 & 0.42 & 2014-10-10.42 & 56940.42 & 7.73 & - & - & - & $5.7\pm1.3$ & $4.3\pm1.0$ \\
00032613051\_5 & 1.05 & 2014-10-10.55 & 56940.55 & 7.86 & - & - & - & $4.1\pm0.7$ & $3.1\pm0.5$ \\
00032613051\_6 & 0.54 & 2014-10-10.62 & 56940.62 & 7.93 & - & - & - & $4.5\pm1.1$ & $3.4\pm0.8$ \\
00032613050\_3 & 0.57 & 2014-10-11.09 & 56941.09 & 8.40 & - & - & $>20.0$ & $1.2\pm0.5$ & $0.9\pm0.4$ \\
00032613050\_4 & 0.81 & 2014-10-11.15 & 56941.15 & 8.46 & - & - & $>20.2$ & $3.0\pm0.7$ & $2.3\pm0.5$ \\
00032613050\_5 & 0.76 & 2014-10-11.22 & 56941.22 & 8.53 & - & - & $>20.1$ & $2.0\pm0.6$ & $1.5\pm0.4$ \\
00032613050\_6 & 0.76 & 2014-10-11.30 & 56941.30 & 8.61 & - & - & $>20.2$ & $2.3\pm0.6$ & $1.8\pm0.5$ \\
\hline
\end{tabular}
\end{center}
\noindent
Notes:\hspace{0.1cm} As for Table\,\ref{tab:obs}.\\
\end{table*}

\clearpage

%
\begin{table*}[ht]
\addtocounter{table}{-1}
\caption{continued.}
\begin{center}
\begin{tabular}{rrrrrrrrrrrr}\hline\hline \noalign{\smallskip}
ObsID\_part & Exp$^a$ & Date$^b$ & MJD$^b$ & $\Delta t^c$ & \multicolumn{3}{c}{UV$^d$ [mag]} & Rate &  L$_{0.2-1.0}$ $^e$\\
& [ks] & [UT] & [d] & [d] & uvm2 & uvw1 & uvw2 & [\power{-2} ct s$^{-1}$] & \power{38} erg s$^{-1}$]\\ \hline \noalign{\smallskip}
00032613050\_7 & 1.12 & 2014-10-11.36 & 56941.36 & 8.67 & - & - & $>20.4$ & $4.7\pm0.7$ & $3.5\pm0.5$ \\
00032613050\_8 & 0.71 & 2014-10-11.55 & 56941.55 & 8.86 & - & - & $20.0\pm0.4$ & $2.8\pm0.7$ & $2.1\pm0.5$ \\
00032613050\_9 & 0.73 & 2014-10-11.63 & 56941.62 & 8.93 & - & - & $>20.2$ & $3.3\pm0.8$ & $2.5\pm0.6$ \\
00032613052\_1 & 0.23 & 2014-10-12.04 & 56942.04 & 9.35 & $>18.9$ & - & - & $5.6\pm1.9$ & $4.2\pm1.4$ \\
00032613052\_2 & 0.50 & 2014-10-12.23 & 56942.23 & 9.54 & $>19.6$ & - & - & $4.1\pm1.0$ & $3.1\pm0.8$ \\
00032613052\_3 & 0.27 & 2014-10-12.44 & 56942.44 & 9.75 & $>19.2$ & - & - & $5.9\pm1.6$ & $4.4\pm1.2$ \\
00032613052\_4 & 1.72 & 2014-10-12.49 & 56942.48 & 9.79 & $>20.3$ & - & - & $4.7\pm0.6$ & $3.5\pm0.5$ \\
00032613052\_5 & 1.72 & 2014-10-12.55 & 56942.55 & 9.86 & $>20.4$ & - & - & $1.7\pm0.4$ & $1.3\pm0.3$ \\
00032613052\_6 & 1.66 & 2014-10-12.62 & 56942.62 & 9.93 & $>20.4$ & - & - & $5.1\pm0.6$ & $3.8\pm0.5$ \\
00032613052\_7 & 0.59 & 2014-10-12.83 & 56942.83 & 10.14 & $>19.8$ & - & - & $4.4\pm1.0$ & $3.3\pm0.8$ \\
00032613052\_8 & 1.04 & 2014-10-13.55 & 56943.55 & 10.86 & - & $>20.1$ & - & $2.4\pm0.6$ & $1.8\pm0.4$ \\
00032613054\_1 & 0.40 & 2014-10-13.57 & 56943.57 & 10.88 & - & $>19.5$ & - & $4.2\pm1.2$ & $3.2\pm0.9$ \\
00032613054\_2 & 1.43 & 2014-10-13.62 & 56943.62 & 10.93 & - & $>20.2$ & - & $4.0\pm0.6$ & $3.0\pm0.4$ \\
00032613055\_1 & 0.31 & 2014-10-14.04 & 56944.04 & 11.35 & - & - & - & $0.9\pm0.6$ & $0.6\pm0.5$ \\
00032613055\_2 & 0.21 & 2014-10-14.30 & 56944.30 & 11.61 & - & - & - & $3.3\pm1.4$ & $2.5\pm1.1$ \\
00032613055\_3 & 0.70 & 2014-10-14.37 & 56944.37 & 11.68 & - & - & - & $3.3\pm0.8$ & $2.5\pm0.6$ \\
00032613055\_4 & 0.64 & 2014-10-14.43 & 56944.43 & 11.74 & - & - & - & $5.3\pm1.0$ & $4.0\pm0.8$ \\
00032613055\_5 & 1.74 & 2014-10-14.49 & 56944.48 & 11.79 & - & - & - & $3.5\pm0.5$ & $2.6\pm0.4$ \\
00032613055\_6 & 1.72 & 2014-10-14.55 & 56944.55 & 11.86 & - & - & - & $2.7\pm0.5$ & $2.1\pm0.4$ \\
00032613056\_1 & 0.11 & 2014-10-15.42 & 56945.42 & 12.73 & - & - & $>19.0$ & $<15.2$ & $<11.4$ \\
00032613056\_2 & 0.36 & 2014-10-15.49 & 56945.49 & 12.80 & - & - & $>19.7$ & $4.4\pm1.3$ & $3.3\pm1.0$ \\
00032613056\_3 & 0.47 & 2014-10-15.55 & 56945.55 & 12.86 & - & - & $>19.8$ & $1.9\pm0.7$ & $1.4\pm0.5$ \\
00032613056\_4 & 0.36 & 2014-10-15.69 & 56945.69 & 13.00 & - & - & $>19.6$ & $3.3\pm1.1$ & $2.5\pm0.8$ \\
00032613056\_5 & 0.34 & 2014-10-15.75 & 56945.75 & 13.06 & - & - & $>19.6$ & $5.8\pm1.5$ & $4.3\pm1.1$ \\
00032613057\_1 & 1.26 & 2014-10-17.36 & 56947.36 & 14.67 & - & $>20.1$ & - & $2.5\pm0.5$ & $1.9\pm0.4$ \\
00032613057\_2 & 1.26 & 2014-10-17.43 & 56947.43 & 14.74 & - & $>20.2$ & - & $3.2\pm0.6$ & $2.4\pm0.5$ \\
00032613057\_3 & 1.26 & 2014-10-17.50 & 56947.50 & 14.81 & - & $>20.1$ & - & $3.0\pm0.5$ & $2.3\pm0.4$ \\
00032613057\_4 & 1.05 & 2014-10-17.63 & 56947.63 & 14.94 & - & $>20.1$ & - & $2.5\pm0.6$ & $1.9\pm0.4$ \\
00032613058\_1 & 0.02 & 2014-10-18.76 & 56948.75 & 16.06 & - & - & - & $<40.2$ & $<30.2$ \\
00032613058\_2 & 1.00 & 2014-10-18.83 & 56948.83 & 16.14 & - & - & - & $1.8\pm0.5$ & $1.3\pm0.4$ \\
00032613058\_3 & 0.20 & 2014-10-18.91 & 56948.91 & 16.22 & - & - & - & $1.9\pm1.2$ & $1.4\pm0.9$ \\
00032613058\_4 & 0.87 & 2014-10-18.97 & 56948.96 & 16.27 & - & - & - & $2.3\pm0.6$ & $1.7\pm0.4$ \\
00032613059\_1 & 0.04 & 2014-10-19.31 & 56949.31 & 16.62 & - & - & $>18.3$ & $2.8\pm2.8$ & $2.1\pm2.1$ \\
00032613059\_2 & 1.08 & 2014-10-19.36 & 56949.36 & 16.67 & - & - & $>20.3$ & $0.7\pm0.3$ & $0.6\pm0.2$ \\
00032613059\_3 & 1.02 & 2014-10-19.43 & 56949.43 & 16.74 & - & - & $>20.3$ & $1.2\pm0.4$ & $0.9\pm0.3$ \\
00032613059\_4 & 1.02 & 2014-10-19.50 & 56949.50 & 16.81 & - & - & $>20.2$ & $1.5\pm0.4$ & $1.1\pm0.3$ \\
00032613059\_5 & 0.99 & 2014-10-19.97 & 56949.96 & 17.27 & - & - & $>20.3$ & $<1.3$ & $<1.0$ \\
00032613060\_1 & 0.23 & 2014-10-20.03 & 56950.04 & 17.35 & $>19.2$ & - & - & $<3.8$ & $<2.9$ \\
00032613060\_2 & 0.94 & 2014-10-20.37 & 56950.37 & 17.68 & $>20.0$ & - & - & $0.5\pm0.3$ & $0.4\pm0.2$ \\
00032613060\_3 & 0.55 & 2014-10-20.57 & 56950.57 & 17.88 & $>19.7$ & - & - & $<2.3$ & $<1.7$ \\
00032613060\_4 & 1.55 & 2014-10-20.63 & 56950.62 & 17.93 & $>20.3$ & - & - & $<1.1$ & $<0.8$ \\
00032613060\_5 & 1.48 & 2014-10-20.83 & 56950.82 & 18.13 & $>20.3$ & - & - & $0.5\pm0.2$ & $0.3\pm0.2$ \\
00032613060\_6 & 0.00 & 2014-10-20.89 & 56950.89 & 18.20 & $>15.5$ & - & - & $<373.0$ & $<280.0$ \\
00032613061\_1 & 0.86 & 2014-10-21.42 & 56951.42 & 18.73 & - & $>20.0$ & - & $<1.4$ & $<1.0$ \\
00032613061\_2 & 0.15 & 2014-10-21.62 & 56951.62 & 18.93 & - & $>20.4$ & - & $<5.1$ & $<3.8$ \\
00032613061\_3 & 0.03 & 2014-10-21.62 & 56951.62 & 18.93 & - & - & - & $<28.6$ & $<21.5$ \\
00032613061\_4 & 1.18 & 2014-10-21.62 & 56951.62 & 18.93 & - & - & - & $<1.4$ & $<1.0$ \\
00032613061\_5 & 0.22 & 2014-10-21.64 & 56951.64 & 18.95 & - & - & - & $<3.6$ & $<2.7$ \\
00032613061\_6 & 1.65 & 2014-10-21.69 & 56951.69 & 19.00 & - & $>20.4$ & - & $<0.6$ & $<0.4$ \\
00032613061\_7 & 1.74 & 2014-10-21.75 & 56951.75 & 19.06 & - & $>20.3$ & - & $<0.6$ & $<0.5$ \\
00032613061\_8 & 0.03 & 2014-10-21.82 & 56951.82 & 19.13 & - & $>17.7$ & - & $<38.4$ & $<28.8$ \\
00032613062\_1 & 0.94 & 2014-10-22.03 & 56952.03 & 19.34 & - & - & - & $<0.8$ & $<0.6$ \\
00032613062\_2 & 0.21 & 2014-10-22.49 & 56952.49 & 19.80 & - & - & - & $<5.7$ & $<4.3$ \\
00032613062\_3 & 1.14 & 2014-10-22.56 & 56952.56 & 19.87 & - & - & - & $<0.9$ & $<0.7$ \\
00032613062\_4 & 0.04 & 2014-10-22.75 & 56952.75 & 20.06 & - & - & - & $<18.2$ & $<13.7$ \\
00032613062\_5 & 0.03 & 2014-10-22.76 & 56952.76 & 20.07 & - & - & - & $<33.2$ & $<24.9$ \\
00032613062\_6 & 0.24 & 2014-10-22.76 & 56952.76 & 20.07 & - & - & - & $<3.5$ & $<2.6$ \\
00032613062\_7 & 0.30 & 2014-10-22.76 & 56952.76 & 20.07 & - & - & - & $<2.7$ & $<2.0$ \\
00032613062\_8 & 0.24 & 2014-10-22.77 & 56952.77 & 20.08 & - & - & - & $<3.5$ & $<2.7$ \\
00032613062\_9 & 0.08 & 2014-10-22.77 & 56952.77 & 20.08 & - & - & - & $<10.7$ & $<8.0$ \\
\hline
\end{tabular}
\end{center}
\noindent
\end{table*}

\clearpage

%
\begin{table*}[ht]
\caption{Individual \swift snapshots of the 2013 monitoring observations from \hndk.}
\label{tab:obs_split13}
\begin{center}
\begin{tabular}{rrrrrrrrrrrr}\hline\hline \noalign{\smallskip}
ObsID\_part & Exp$^a$ & Date$^b$ & MJD$^b$ & $\Delta t^c$ & \multicolumn{3}{c}{UV$^d$ [mag]} & Rate &  L$_{0.2-1.0}$ $^e$\\
& [ks] & [UT] & [d] & [d] & uvm2 & uvw1 & uvw2 & [\power{-2} ct s$^{-1}$] & \power{38} erg s$^{-1}$]\\ \hline \noalign{\smallskip}
00032613005\_1 & 0.7 & 2013-12-03.03 & 56629.03 & 6.43 & - & - & $>20.0$ & $2.6\pm0.7$ & $2.0\pm0.5$ \\
00032613005\_2 & 1.3 & 2013-12-03.09 & 56629.09 & 6.49 & - & - & $19.9\pm0.3$ & $0.9\pm0.3$ & $0.7\pm0.2$ \\
00032613005\_3 & 0.7 & 2013-12-03.16 & 56629.16 & 6.56 & - & - & $>20.0$ & $2.9\pm0.7$ & $2.2\pm0.5$ \\
00032613005\_4 & 1.7 & 2013-12-03.23 & 56629.23 & 6.63 & - & - & $19.9\pm0.2$ & $1.4\pm0.3$ & $1.1\pm0.2$ \\
00032613005\_5 & 1.7 & 2013-12-03.43 & 56629.43 & 6.83 & - & - & $19.2\pm0.2$ & $<1.0$ & $<0.7$ \\
00032613006\_1 & 0.2 & 2013-12-04.09 & 56630.09 & 7.49 & - & - & $>19.3$ & $2.9\pm1.2$ & $2.1\pm0.9$ \\
00032613006\_2 & 1.7 & 2013-12-04.16 & 56630.16 & 7.56 & $>19.9$ & $>19.1$ & $>20.0$ & $1.7\pm0.4$ & $1.3\pm0.3$ \\
00032613006\_3 & 0.0 & 2013-12-04.23 & 56630.23 & 7.63 & - & - & $>18.3$ & $<18.5$ & $<13.9$ \\
00032613007\_1 & 1.9 & 2013-12-04.70 & 56630.70 & 8.10 & $19.3\pm0.3$ & $>19.8$ & $19.8\pm0.3$ & $1.6\pm0.3$ & $1.2\pm0.2$ \\
00032613007\_2 & 0.1 & 2013-12-04.76 & 56630.76 & 8.16 & - & - & $>18.7$ & $3.5\pm2.1$ & $2.6\pm1.6$ \\
00032613008\_1 & 1.7 & 2013-12-05.23 & 56631.23 & 8.63 & $>19.7$ & $>19.8$ & $>19.9$ & $2.2\pm0.4$ & $1.7\pm0.3$ \\
00032613009\_1 & 1.3 & 2013-12-05.70 & 56631.70 & 9.10 & $>19.5$ & $>19.6$ & $>19.8$ & $4.2\pm0.6$ & $3.1\pm0.5$ \\
00032613009\_2 & 0.3 & 2013-12-05.85 & 56631.85 & 9.25 & $>18.8$ & $>17.7$ & $>19.1$ & $4.1\pm1.4$ & $3.1\pm1.1$ \\
00032613009\_3 & 0.3 & 2013-12-05.96 & 56631.96 & 9.36 & $>18.4$ & - & $>18.8$ & $4.3\pm1.3$ & $3.2\pm1.0$ \\
00032613010\_1 & 1.0 & 2013-12-06.10 & 56632.10 & 9.50 & $>19.3$ & $>19.2$ & $>19.4$ & $3.7\pm0.7$ & $2.8\pm0.5$ \\
00032613010\_2 & 1.0 & 2013-12-06.16 & 56632.16 & 9.56 & $>19.4$ & $>19.2$ & $>19.5$ & $3.8\pm0.7$ & $2.8\pm0.5$ \\
00032613011\_1 & 1.5 & 2013-12-06.63 & 56632.63 & 10.03 & $>19.6$ & $>19.7$ & $>19.8$ & $4.0\pm0.6$ & $3.0\pm0.4$ \\
00032613011\_2 & 0.5 & 2013-12-06.71 & 56632.71 & 10.11 & $>19.0$ & $>19.0$ & $>19.2$ & $4.2\pm1.0$ & $3.2\pm0.7$ \\
00032613012\_5 & 0.0 & 2013-12-07.10 & 56633.10 & 10.50 & - & $>18.8$ & $>19.6$ & $9.6\pm6.8$ & $7.2\pm5.1$ \\
00032613012\_2 & 0.0 & 2013-12-07.10 & 56633.10 & 10.50 & - & - & - & $4.4\pm4.4$ & $3.3\pm3.3$ \\
00032613012\_3 & 0.0 & 2013-12-07.10 & 56633.10 & 10.50 & - & - & - & $16.0\pm9.2$ & $12.0\pm6.9$ \\
00032613012\_4 & 0.0 & 2013-12-07.10 & 56633.10 & 10.50 & - & - & - & $8.7\pm6.2$ & $6.5\pm4.7$ \\
00032613012\_1 & 0.0 & 2013-12-07.10 & 56633.10 & 10.50 & - & - & - & $4.8\pm4.8$ & $3.6\pm3.6$ \\
00032613012\_6 & 0.0 & 2013-12-07.10 & 56633.11 & 10.51 & - & - & - & $<79.2$ & $<59.4$ \\
00032613012\_7 & 0.2 & 2013-12-07.11 & 56633.11 & 10.51 & $>18.9$ & - & - & $<9.6$ & $<7.2$ \\
00032613012\_8 & 1.4 & 2013-12-07.17 & 56633.16 & 10.56 & $>19.8$ & $>18.8$ & $>19.6$ & $1.3\pm0.4$ & $1.0\pm0.3$ \\
00032613012\_9 & 0.2 & 2013-12-07.23 & 56633.23 & 10.63 & $>20.0$ & $>18.8$ & $>19.6$ & $1.3\pm0.9$ & $0.9\pm0.7$ \\
00032613013\_1 & 1.1 & 2013-12-07.57 & 56633.57 & 10.97 & $>19.3$ & $>19.3$ & $>19.6$ & $1.7\pm0.4$ & $1.3\pm0.3$ \\
00032613013\_2 & 0.0 & 2013-12-07.63 & 56633.63 & 11.03 & - & - & $>19.4$ & $23.9\pm24.0$ & $18.0\pm18.0$ \\
00032613013\_3 & 0.3 & 2013-12-07.63 & 56633.63 & 11.03 & - & - & - & $2.0\pm0.9$ & $1.5\pm0.6$ \\
00032613013\_4 & 0.1 & 2013-12-07.64 & 56633.64 & 11.04 & $>19.2$ & - & - & $<15.1$ & $<11.3$ \\
00032613013\_6 & 0.3 & 2013-12-07.64 & 56633.64 & 11.04 & - & $>19.1$ & - & $2.4\pm1.0$ & $1.8\pm0.8$ \\
00032613013\_5 & 0.0 & 2013-12-07.64 & 56633.64 & 11.04 & - & - & - & $<106.0$ & $<79.6$ \\
00032613013\_7 & 0.1 & 2013-12-07.64 & 56633.64 & 11.04 & - & - & - & $<15.1$ & $<11.3$ \\
00032613014\_1 & 0.2 & 2013-12-08.03 & 56634.04 & 11.44 & - & - & $>19.1$ & $4.6\pm1.8$ & $3.4\pm1.4$ \\
00032613014\_2 & 1.4 & 2013-12-08.17 & 56634.17 & 11.57 & $>19.5$ & $>19.5$ & $>19.8$ & $2.5\pm0.5$ & $1.9\pm0.4$ \\
00032613015\_1 & 1.1 & 2013-12-08.57 & 56634.57 & 11.97 & $>19.2$ & $>19.5$ & $>19.6$ & $3.5\pm0.6$ & $2.6\pm0.5$ \\
00032613015\_2 & 0.8 & 2013-12-08.64 & 56634.64 & 12.04 & $>19.0$ & $>19.2$ & $>19.4$ & $2.6\pm0.7$ & $1.9\pm0.5$ \\
00032613016\_1 & 0.1 & 2013-12-09.04 & 56635.04 & 12.44 & - & $>18.2$ & - & $3.6\pm2.5$ & $2.7\pm1.9$ \\
00032613016\_2 & 0.1 & 2013-12-09.10 & 56635.10 & 12.50 & - & $>18.2$ & - & $4.2\pm2.9$ & $3.1\pm2.2$ \\
00032613016\_3 & 0.1 & 2013-12-09.17 & 56635.17 & 12.57 & - & $>18.2$ & - & $<11.2$ & $<8.4$ \\
00032613016\_4 & 0.4 & 2013-12-09.17 & 56635.17 & 12.57 & - & $>18.8$ & $>19.2$ & $2.1\pm0.9$ & $1.5\pm0.7$ \\
00032613016\_5 & 1.6 & 2013-12-09.50 & 56635.50 & 12.90 & $>19.8$ & $>19.2$ & $>19.8$ & $2.9\pm0.5$ & $2.1\pm0.4$ \\
00032613016\_6 & 1.7 & 2013-12-09.57 & 56635.57 & 12.97 & $>19.8$ & $>19.1$ & $>19.7$ & $2.0\pm0.4$ & $1.5\pm0.3$ \\
00032613016\_7 & 0.5 & 2013-12-09.64 & 56635.64 & 13.04 & $>18.9$ & $>18.4$ & $>19.0$ & $3.7\pm0.9$ & $2.8\pm0.7$ \\
00032613017\_1 & 1.7 & 2013-12-10.44 & 56636.44 & 13.84 & $>19.7$ & $>19.8$ & $>19.9$ & $3.2\pm0.5$ & $2.4\pm0.4$ \\
00032613017\_2 & 1.7 & 2013-12-10.50 & 56636.50 & 13.90 & $>19.6$ & $>19.8$ & $>19.8$ & $3.5\pm0.5$ & $2.6\pm0.4$ \\
00032613017\_3 & 0.4 & 2013-12-10.57 & 56636.57 & 13.97 & $>18.6$ & $>18.8$ & $>18.9$ & $2.5\pm0.9$ & $1.9\pm0.7$ \\
00032613018\_1 & 0.4 & 2013-12-11.04 & 56637.04 & 14.44 & $>18.6$ & $>18.7$ & $>18.8$ & $<3.0$ & $<2.3$ \\
00032613018\_2 & 0.8 & 2013-12-11.11 & 56637.11 & 14.51 & $>19.2$ & $>19.3$ & $>19.4$ & $2.4\pm0.6$ & $1.8\pm0.5$ \\
00032613018\_3 & 0.9 & 2013-12-11.17 & 56637.17 & 14.57 & $>19.2$ & $>19.3$ & $>19.5$ & $1.5\pm0.5$ & $1.1\pm0.4$ \\
00032613019\_1 & 0.3 & 2013-12-11.31 & 56637.30 & 14.70 & $>18.7$ & $>18.0$ & $>18.5$ & $2.1\pm1.0$ & $1.6\pm0.8$ \\
00032613019\_2 & 1.7 & 2013-12-11.50 & 56637.50 & 14.90 & $>19.8$ & $>19.1$ & $>19.8$ & $1.4\pm0.3$ & $1.0\pm0.3$ \\
\hline
\end{tabular}
\end{center}
\noindent
Notes:\hspace{0.1cm} As for Table\,\ref{tab:obs} except for $^c $ which gives the time in days after the 2013 outburst of \nova in the optical on 2013-11-26.60 UT (MJD = 56622.60), following \hndk.\\
\end{table*}

\clearpage

%
\begin{table*}[ht]
\addtocounter{table}{-1}
\caption{continued.}
\begin{center}
\begin{tabular}{rrrrrrrrrrrr}\hline\hline \noalign{\smallskip}
ObsID\_part & Exp$^a$ & Date$^b$ & MJD$^b$ & $\Delta t^c$ & \multicolumn{3}{c}{UV$^d$ [mag]} & Rate &  L$_{0.2-1.0}$ $^e$\\
& [ks] & [UT] & [d] & [d] & uvm2 & uvw1 & uvw2 & [\power{-2} ct s$^{-1}$] & \power{38} erg s$^{-1}$]\\ \hline \noalign{\smallskip}
00032613020\_1 & 1.5 & 2013-12-11.57 & 56637.57 & 14.97 & $>19.6$ & $>19.7$ & $>19.8$ & $1.9\pm0.4$ & $1.4\pm0.3$ \\
00032613020\_2 & 0.3 & 2013-12-11.64 & 56637.64 & 15.04 & $>18.5$ & $>18.6$ & $>18.8$ & $1.4\pm0.7$ & $1.1\pm0.6$ \\
00032613019\_3 & 1.0 & 2013-12-11.64 & 56637.64 & 15.04 & $>19.6$ & $>18.9$ & $>19.5$ & $1.9\pm0.5$ & $1.5\pm0.4$ \\
00032613021\_1 & 0.2 & 2013-12-12.04 & 56638.04 & 15.44 & $>18.9$ & - & - & $1.6\pm1.2$ & $1.2\pm0.9$ \\
00032613021\_2 & 0.9 & 2013-12-12.11 & 56638.11 & 15.51 & $19.8\pm0.4$ & - & - & $1.1\pm0.4$ & $0.8\pm0.3$ \\
00032613021\_3 & 0.8 & 2013-12-12.17 & 56638.17 & 15.57 & $>19.9$ & - & - & $1.3\pm0.4$ & $0.9\pm0.3$ \\
00032613022\_1 & 0.7 & 2013-12-12.32 & 56638.32 & 15.72 & $>19.8$ & - & - & $1.9\pm0.5$ & $1.4\pm0.4$ \\
00032613022\_2 & 0.7 & 2013-12-12.39 & 56638.39 & 15.79 & $>19.8$ & - & - & $1.2\pm0.4$ & $0.9\pm0.3$ \\
00032613022\_3 & 0.4 & 2013-12-12.45 & 56638.45 & 15.85 & $>19.6$ & - & - & $<3.1$ & $<2.3$ \\
00032613023\_1 & 1.2 & 2013-12-12.52 & 56638.52 & 15.92 & $>19.6$ & $>19.5$ & $>19.8$ & $2.2\pm0.5$ & $1.6\pm0.4$ \\
00032613023\_2 & 0.7 & 2013-12-12.58 & 56638.58 & 15.98 & $>19.2$ & $>19.1$ & $>19.4$ & $1.2\pm0.4$ & $0.9\pm0.3$ \\
00032613024\_1 & 0.7 & 2013-12-12.78 & 56638.78 & 16.18 & $>19.8$ & - & - & $1.5\pm0.5$ & $1.1\pm0.4$ \\
00032613024\_2 & 0.6 & 2013-12-12.85 & 56638.85 & 16.25 & $>19.8$ & - & - & $1.3\pm0.5$ & $0.9\pm0.4$ \\
00032613024\_3 & 0.8 & 2013-12-12.92 & 56638.92 & 16.32 & $>20.0$ & - & - & $1.5\pm0.5$ & $1.1\pm0.4$ \\
00032613025\_1 & 1.1 & 2013-12-13.11 & 56639.11 & 16.51 & - & $>20.1$ & - & $0.7\pm0.3$ & $0.5\pm0.2$ \\
00032613025\_2 & 0.9 & 2013-12-13.17 & 56639.17 & 16.57 & - & $>19.9$ & - & $1.3\pm0.4$ & $1.0\pm0.3$ \\
00032613026\_1 & 1.1 & 2013-12-13.31 & 56639.31 & 16.71 & - & $>20.1$ & - & $0.7\pm0.3$ & $0.5\pm0.2$ \\
00032613026\_2 & 0.8 & 2013-12-13.38 & 56639.38 & 16.78 & - & $>20.0$ & - & $<2.0$ & $<1.5$ \\
00032613027\_1 & 1.1 & 2013-12-13.58 & 56639.58 & 16.98 & - & $>20.1$ & - & $0.5\pm0.2$ & $0.4\pm0.2$ \\
00032613027\_2 & 0.8 & 2013-12-13.65 & 56639.65 & 17.05 & - & $>20.0$ & - & $0.5\pm0.3$ & $0.4\pm0.2$ \\
00032613028\_1 & 1.1 & 2013-12-13.78 & 56639.78 & 17.18 & $>19.4$ & $>19.4$ & $>19.6$ & $<1.4$ & $<1.0$ \\
00032613028\_2 & 0.9 & 2013-12-13.85 & 56639.85 & 17.25 & $>19.3$ & $>19.4$ & $>19.6$ & $<1.4$ & $<1.0$ \\
00032613029\_1 & 0.3 & 2013-12-14.24 & 56640.25 & 17.65 & - & - & - & $<3.2$ & $<2.4$ \\
00032613030\_1 & 0.9 & 2013-12-14.38 & 56640.38 & 17.78 & - & - & - & $0.8\pm0.3$ & $0.6\pm0.3$ \\
00032613031\_1 & 1.1 & 2013-12-14.52 & 56640.52 & 17.92 & - & - & - & $<0.9$ & $<0.7$ \\
00032613031\_2 & 0.3 & 2013-12-14.60 & 56640.59 & 17.99 & - & - & - & $<2.8$ & $<2.1$ \\
00032613031\_3 & 0.3 & 2013-12-14.66 & 56640.66 & 18.06 & - & - & - & $<4.2$ & $<3.2$ \\
00032613031\_4 & 0.2 & 2013-12-14.71 & 56640.71 & 18.11 & - & - & - & $<3.7$ & $<2.8$ \\
00032613032\_1 & 0.8 & 2013-12-14.78 & 56640.78 & 18.18 & $>19.2$ & $>19.3$ & $>19.5$ & $<2.1$ & $<1.5$ \\
00032613032\_2 & 1.0 & 2013-12-14.85 & 56640.85 & 18.25 & $>19.5$ & $>19.5$ & $>19.7$ & $<1.2$ & $<0.9$ \\
\hline
\end{tabular}
\end{center}
\noindent
\end{table*}

\end{document}